\documentclass[twocolumn,twoside,slac]{revtex4}

\usepackage{graphicx}
\usepackage{makeidx}
\usepackage{fancyhdr}

\begin{document}

\title{Novel Data Acquisition System for Silicon Tracking Detectors}

\author{L.~A.~Wendland, K.~Banzuzi, S.~Czellar, A.~Heikkinen, J.~H\"ark\"onen,
P.~Johansson, V.~Karim\"aki, T.~Lamp\'en, P.~Luukka, P.~Meht\"al\"a,
J.~Niku, S.~Nummela, J.~Nysten, J.~Simpura, E.~Tuovinen, E.~Tuominen,
J.~Tuominiemi, D.~Ungaro, T.~Vaarala, M.~Voutilainen, A.~Zibellini}
\affiliation{Helsinki Institute of Physics, P.O. Box 64, FIN-00014
  University of Helsinki, Finland}

\begin{abstract}
We have developed a novel data acquisition system for measuring
tracking parameters of a silicon detector in a particle beam. The
system is based on a commercial Analog-to-Digital VME module and a
PC Linux based Data Acquisition System.
This DAQ is realized with C++ code
using object-oriented techniques. Track parameters for the beam
particles were reconstructed using off-line analysis code and
automatic detector position alignment algorithm.

The new DAQ was used to test novel Czochralski type silicon
detectors. The important silicon detector
parameters, including signal size distributions and signal to noise
distributions, were successfully extracted from the detector under
study. The efficiency of the detector was measured to be 95 \%, the
resolution about 10 $\mu$m, and the signal to noise ratio about 10.

\end{abstract}

\maketitle

\thispagestyle{fancy}


\section{INTRODUCTION}
In order to cope with the challenging LHC-like running environment of
high event rates, high number of channels, limited processing time and
limited storing capability, the Data Acquisition System (DAQ) of the
Helsinki Silicon Beam Telescope (SiBT)~\cite{SiBT_hardware99, SiBT2000a} was
rebuilt. The SiBT device is operated by Helsinki Institute of Physics
at the CERN H2 beam line. It is used
to measure tracks of incoming particles with high resolution to
offer reference tracks to the CERN CMS subdetector systems for testing
their detectors at the H2 beam. The SiBT is based
on position sensitive silicon detectors attached to appropriate readout
electronics and DAQ.~\cite{daq_tdr, daq_tdr2}

\section{EXPERIMENTAL SETUP}
\begin{figure*}
  \includegraphics[width=135mm,keepaspectratio]{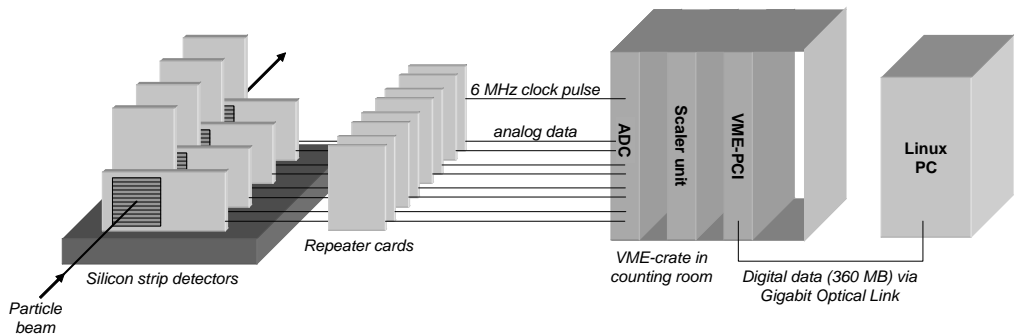}
  \caption{Schematical layout of the SiBT experiment.}
  \label{fig_sibtdaq_setup}
\end{figure*}

\subsection{Silicon strip detectors}
Fig.~\ref{fig_sibtdaq_setup} shows the schematical layout of the SiBT
setup.
The SiBT apparatus consists of eight single-sided silicon microstrip
detector planes. The detectors with 1024 p+ strips are processed on
n-type Float Zone silicon wafers that are 300 $\mu$m thick.
The width of the strips is 14 $\mu$m and pitch 55 $\mu$m giving a total
detector area of 5.6 $\times$ 5.6 cm$^{2}$~\cite{harkonen2002a}. 
Each silicon detector is glued and bonded on a hybrid containing
8~VA1 readout chips.

The eight detectors are installed as four pairs orthogonal to the beam
axis. Every pair consists of one detector with strips positioned
horizontally and another
with strips positioned vertically. The aluminum housings, which
enclose the detector hybrids, are attached on aluminum forks that are
mounted on an optical precision bench consisting of a 492 mm long
brass rail sitting on top of a granite block.

During the summer 2002 tests, a muon beam of the energy of 225 GeV was
used to study the performance of the SiBT and a new Czochralski type
silicon detector.

\subsection{Front-end electronics}
Each silicon detector has a separate front-end electronics unit
consisting of 8 daisy-chained VA1 readout chips and a repeater
card. Each VA1 chip has 128 identical channels with a charge sensitive
amplifier, a shaper and a track-and-hold circuitry in each. The output
of each detector is multiplexed and activated by the output
shift register using a clock signal of 6 MHz.

The readout chips of a hybrid are controlled from the counting room. 
The controls cable has a 50-pin flat cable connector that is connected
to the protective housing to a repeater card mounted 
on an additional rail next to the detector housings. Each repeater card
contains four voltage regulators to supply the readout chips and a
high-speed differential I/O amplifier to buffer and amplify the
balanced analog output signal. Line receivers and analog and digital
circuits control the readout electronics and set the necessary
working parameters for both the detector and the VA1 chips. In
addition, the repeater card has an efficient protection circuit
against breakdown damage in latch-up conditions. All supply voltages
are cut off sharply, if one of the input power lines fail to work.

\subsection{Triggering}
Mounted before the first and after the last detector are two standard
plastic scintillators attached to photomultiplier tubes that are
shielded against a magnetic field of up to 1 T. The photomultiplier
signals are shaped and put through a coincidence unit in the counting
room and the resulting trigger signal is sent back to the repeater
cards to activate the readout modules. The trigger signal starts the
readout sequence simultaneously on each of the eight repeater cards.

Analog delay is added to the trigger signal to achieve optimum charge
collection time in the detectors. The triggering system is protected
against events that are very close to each other by vetoing any
additional trigger signals that should occur while the repeater cards
are busy reading the last event. The dead time for reading one event
using a 6 MHz clock on the repeater cards and taking into account
cable delays is 172 $\mu$s.

\section{DATA ACQUISITION SYSTEM} %

\subsection{Digitization}
The readout of the detectors is controlled from the counting room
using standard commercially available VME-based technology.
A commercial 8-channel VME ADC card SIS3300 with 12-bit resolution is
used to digitize the analog signals from the repeater cards. The eight
repeater cards are connected through equally long coaxial cables to
the ADC. A ninth cable is used to send the clock pulse of one of
the repeater cards to the ADC to trigger the sampling.

The ADC samples simultaneously all the eight channels. It is operated
in asynchronous clock mode, in which the 6 MHz clock pulse of the
repeater cards activates the taking of one sample simultaneously from
all channels at the 100 MHz internal clock rate. Each channel has an
independent ADC circuit and the data of each channel pair is combined
as one 32-bit word and pushed via pipeline into a 128 ksample
SDRAM memory bank. The ADC can thus store 128 events containing data
of the 1024 channels of a single detector.

There are two equal size memory banks on the ADC for each channel pair
to allow operation in circular buffer mode. In circular buffer mode
the active bank is automatically switched to the empty bank once the
active bank has become full and data taking can resume without any
delay and thus no additional dead time is caused.

The ADC provides automatically a time stamp for each event with respect
to the first event in the bank with a time resolution of its internal
clock. The blocks of 128 events are thus internally synchronized with
high precision and therefore, to achieve overall synchronization of
the events, only the first event of the block needs to be synchronized
with the accelerator time frame.

\subsection{Timing}
The DAQ is equipped with time stamping capabilities to ensure
the synchronizability of the events with other experiments using the
same testbeam facility. A fast commercial VME scaler unit is used to
provide time stamps for the first event in the ADC bank. The 100 MHz
internal clock of the ADC is used to provide the count, which is
synchronized to the accelerator hardware by resetting the count at one
second before the spill starts. In the future it is planned to use a
custom-built multi-DAQ card specially designed for high accuracy and
reliable event timing.

\subsection{VME-PCI interface}
The VME crate is controlled with a commercial VME-PCI interface card
SIS3100. It uses a Gigabit Optical Link (GOL) to transmit data via an
optical fiber between the DAQ PC and the VME crate. The interface is
also equipped with 512 MB of SDRAM and a programmable SHARC circuit,
which could be used to process the signals. The optical link has,
however, proven to be sufficiently fast at the current DAQ speeds for
pushing the data directly from the VME-bus into the memory of the
DAQ-PC without storing it to the memory of the interface card.

The interface card also includes three digital input and output
channels, which are used to give information for the DAQ PC about the
phase of the accelerator.

\section{DAQ SOFTWARE} %
Because the accelerator cycle consists
of a spill period, during which particles are extracted to the test
beam area,
and an idle period, during which new particles are accelerated and other
beams are served, it is logical to divide the DAQ operation into two
operation cycles. When particles are passing through the detector and
thus producing data during spill cycle, the DAQ is concentrating only
on reading the new data in order to maximize the number of events that
are read. This raw event data is further processed and stored on a
hard drive during the idle period while waiting for new data.

The key challenge that the structure of the DAQ software faces is the
fact that large amounts of data have to be read and processed reliably
in real time. Usually, the amount of data that needs to be transferred
to a PC is reduced by using pipelined programmable circuits inside the
VME crate. Because no fast enough commercial modules were available, a
custom built VME card or a large number of costly ADC modules would
have been needed. The GOL of the VME-PCI
interface was found to have a high enough data transfer rate to allow
the transfer of all the raw data to the memory of the PC with minimal
performance losses. This way the data processing is more reliable as
the whole process can be monitored on the PC and the physics cuts on
the data may be altered in real time.

\begin{figure*}
  \includegraphics[width=100mm,keepaspectratio]{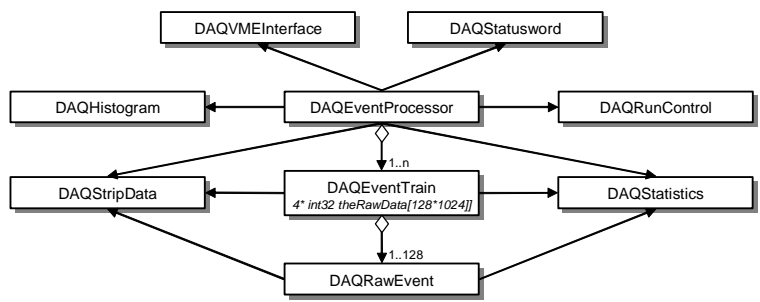}
  \caption{UML-diagram of the DAQ software.}
  \label{fig_sibtdaq_uml}
\end{figure*}

\subsection{Software architecture}
The DAQ software is realized using C++ object orientated techniques on
a PC equipped with a 1200 MHz AMD Athlon processor and running CERN
RedHat Linux 7.1. The UML diagram of the software 
is shown in Fig.~\ref{fig_sibtdaq_uml}. The raw data is stored as
a vector of DAQEventTrain objects each containing four 128K long
arrays of 32-bit unsigned integers and 128 32-bit time stamps thus
providing storage for a total of 128 events. Each DAQEventTrain
object further contains a vector of 128 DAQRawEvent objects. The
DAQRawEvent objects contain pointers to the data that is in the
DAQEventTrain object and the data can thus be accessed from both the
DAQEventTrain and DAQRawEvent objects.

The pedestal and noise calibration data for the individual detector
channels are stored into the DAQStripData object and accessed directly
from the events during processing. The detector channels are labeled
dead or noisy if the noise level is outside certain user-defined
thresholds and the user may also deliberately mark known channels
to be ignored.

The DAQStatistics object contains information on how the selection
process is working, what percentage of channels is passing through
each selection level and what are the efficiencies for having at least
one hit on a detector plane per event. More detailed information about
the quality of data and the behavior of the detector planes is stored
into the DAQHistogram object, which is an interface to the
Histo-Scope histogramming package. The histograms may be viewed
online, but it does significantly slow down the DAQ operation.

Data concerning the operation and status of the DAQ are stored into
DAQRunControl and DAQStatusWord objects. The memory space of
the VME crate is accessed via a C-based driver that is controlled
through the DAQVMEInterface object.

In order to retain real time operation, the memory of the whole data
structure is dynamically allocated from the memory of the PC as the 
software is started. Thus no time consuming memory allocation needs to
be conducted during the operation of the software. The drawback is
that large amounts of memory are required, as a spill cycle length of
4.8 seconds requires about 360 MB of memory for the raw data.

\subsection{Reading data from hardware}
\begin{figure*}
  \includegraphics[width=135mm,keepaspectratio]{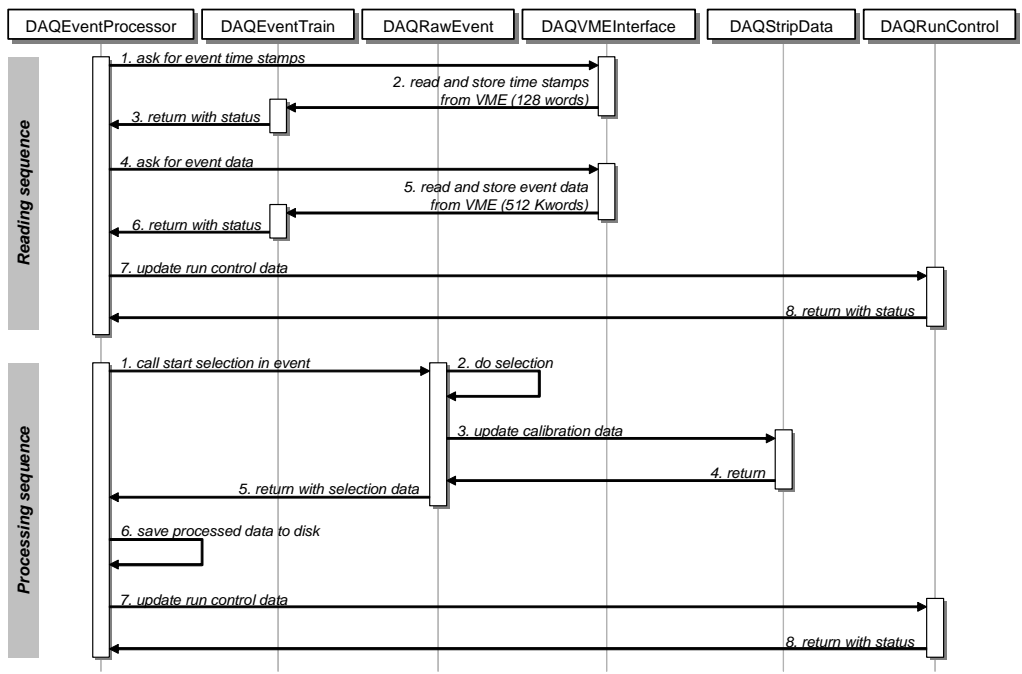}
  \caption{Flow diagram of the reading and data processing sequence of
  the DAQ software.}
  \label{fig_sibtdaq_flow}
\end{figure*}

Once the accelerator hardware provides a pre-spill signal one second
before the spill starts, the software prepares to start reading the
data and arms the hardware. Once the spill is on, the software keeps
checking, whether an ADC bank has been filled with data. If an ADC
bank is full, the software reads the sample data, which has been
stored into four memory banks, each containing the data of two channels.

Fig.~\ref{fig_sibtdaq_flow} shows the flow of the reading sequence.
The data is read to the actual DAQEventTrain object in four blocks
of 128 ksamples of 32-bit data via the VME-PCI interface using
2eVME cycles. Previous data is overwritten. Additionally, the
ADC-internal time stamp register of the events is read. The speed of
the reading process is optimized by minimizing the number of blocks to
be read, since in this way the number of time-costly interrupts on the
VME bus can be minimized.

Once the reading of the blocks is complete, the bank full flag of the
ADC is cleared and the software waits for the first event to arrive in
order to catch its time stamp. The reading sequence repeats this
pattern until the spill is over. To maximize the number of events that
can be read, the software does not answer any user commands during the
reading sequence.

\subsection{Data processing}
The data structure of the software allows the data to be accessed
either on a
single event basis or as blocks of 128 events. This is crucial for the
performance of the software, because this structure allows the
software optimization of both reading and processing the data. The
reading of the data is at its fastest, if the data can be read in as
large blocks as possible from the VME-bus thus minimizing the number
of time consuming VME interrupts. On the other hand, if the events can
be accessed on a single event basis during processing, several
if-clauses are rendered obsolete and valuable processing time is
saved.

In order to process the data efficiently without compromising the
physical performance, three different selection levels are imposed on
the data. On the average, 99.6 \% of the raw data is just electronical
noise, which has to be filtered out of the data of maximally
$1.9 \cdot 10^{8}$ samples (360 MB) during 10.9 seconds. In order to
complete this task, three selection levels are imposed on the data.

For each strip $i$ of each detector a long term average pedestal $P$
is modeled using an IIR filter described by
\begin{equation}
P_{i}^{n} = \frac{m-1}{m} P_{i}^{n-1} + \frac{1}{m}R_{i}^{n},
\label{eq_sibtdaq_pedestal}
\end{equation}
where $m=64$, $R$ the raw data value and $n$ is the current
event. Additionally, the noise $N$ of each strip approximating the
standard deviation is calculated with
\begin{equation}
N_{i}^{n} = \frac{m-1}{m} N_{i}^{n-1} + \frac{1}{m}|R_{i}^{n}-P_{i}^{n}|,
\label{eq_sibtdaq_noise}
\end{equation}
where $m=32$. 

In the first selection level, the raw data value is compared to its
corresponding old pedestal value
\begin{equation}
R - P > 0.
\end{equation}
If this criterion isn't matched, the strip is discarded. With this
simple and fast rejection 50 \% of the noise is rejected. Only if the
criterion is passed, the new pedestal and noise values are calculated
for the strip. This method can be safely used, because the noise
distribution of the pulse height spectrum is symmetrical around zero.

The second selection level enhances the first selection level by
demanding, that a certain constant level of pedestal $k_{P}$
individually specifiable for each detector plane is exceeded
\begin{equation}
R - P > k_{P}.
\label{eq_level2}
\end{equation}
This criterion can be used to reject a further 80-90 \% of the
remaining noise and thus additional processing time is saved before a
more elaborate and exact cut is made on the data.

Once the second selection level has been passed, the signal $S$ is
calculated by subtracting the noise and pedestal from the raw data:
\begin{equation}
S = R - P - N
\end{equation}
The third selection level is defined as
\begin{equation}
S > (N + k_{N_{1}}) \cdot k_{N_{2}},
\end{equation}
where $k_{N_{1}}$ is a small constant set to avoid dead strips and
$k_{N_{2}}$ is usually set to 4.5 in order to make approximately a 4.5
$\sigma$ cut on the signal.

If the third selection level is passed, the actual strip and its
neighboring left and right strips are selected to make a cluster and
they are saved with all their data values to disk. The events are
attached with header information describing the run, spill and event
numbers as well as the time stamp to allow offline synchronization
with data of other experiments. The size of saved data during one
spill cycle amounts on the average to 6.5 MB for 22800 events
including the  event header information. The flow of the processing
sequence is shown in Fig.~\ref{fig_sibtdaq_flow}.

\subsection{User interface}
The DAQ software is operated via a graphical user interface (GUI) that
has been created with C++ using the Qt package~\cite{qt}. The GUI has
direct access via pointers to all the objects of the DAQ software and
thus it doesn't need any additional objects to propagate the data. The
GUI is updated and its requests are executed only when the program is
not busy. Therefore, during the reading and processing periods, the
GUI is neither updated nor responding to user actions in order to
provide the maximum amount of CPU resources to the more important
tasks. User actions are executed in the idle time after the processing
has been finished and while the DAQ is waiting for the next spill.

The user may conveniently define data collection and processing
parameters through the GUI as well as monitor the performance of the
separate detectors and individual detector channels. A light histogram
package Histo-Scope is integrated into the GUI to allow monitoring of
the detector hardware and data quality. The histogrammer is however
intended for slow data taking, as it will considerably slow down the
processing.

\section{RESULTS OF CZOCHRALSKI SETUP}

\begin{figure}
  \includegraphics[width=80mm,keepaspectratio]{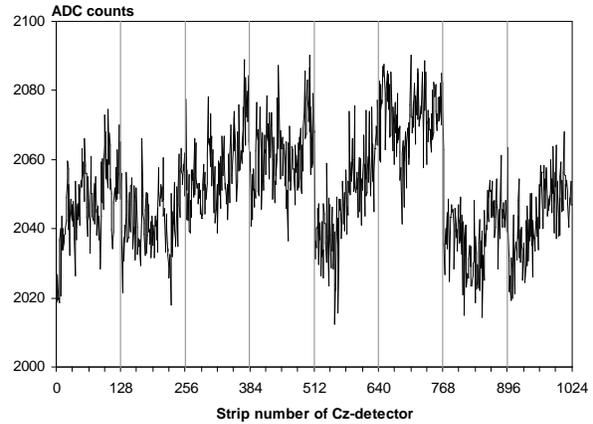}%
  \caption{Pedestal distribution for the individual channels of the
    Cz-detector calculated with Eq.~\ref{eq_sibtdaq_pedestal}. The
    discrete jumps at the edge of each 128 channels are caused by
    different levels in the readout chips.}
  \label{fig_sibtdaq_czped}
\end{figure}

\begin{figure}
  \includegraphics[width=80mm,keepaspectratio]{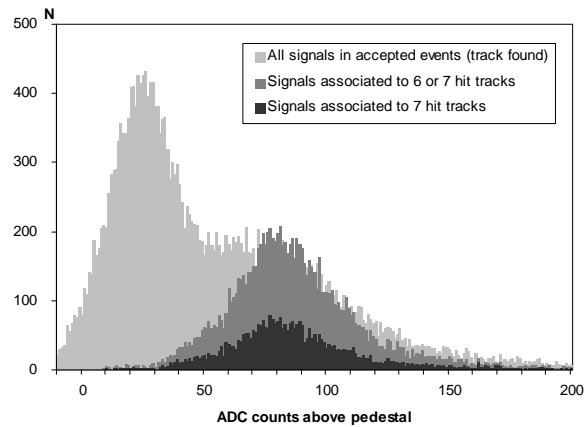}%
  \caption{Pulse height distribution of Cz-detector. The light gray
    color shows the pulse height distribution for the data that has
    passed through selection level of Eq. \ref{eq_level2} with a small
    $k_{P}$. The darker shades of gray show the distribution of the
    pulses for data that has been reconstructed to belong to
    tracks.}
  \label{fig_sibtdaq_selcz}
\end{figure}

The modeling of the detector channel properties with a fast algorithm
is essential, because the channels have to be continuously
calibrated. Fig.~\ref{fig_sibtdaq_czped} shows that the pedestal
modeling is functioning well for the detector channels. It is
noticeable that there are discrete transitions after every 128
nels, because of the different noise levels in the VA1 amplifier
chips on the detectors.

The noise filtering was found to work excellently using the three-step
selection levels. Fig.~\ref{fig_sibtdaq_selcz} shows the pulse height
distribution recorded for the Cz-detector. It displays in arbitrary units
how the signal (dark gray tones) is cleanly extracted using the 
cut of the selection levels and the offline track fitting parameters
from the total data (light gray). Most of the noise is already cut 
away before the figure and thus the noise doesn't peak around zero ADC
counts and the difference between 6 and 7 hits is accounted for one
faulty detector while one detector was completely taken away from
the setup. The plot shows that it is possible to use a fast algorithm
based on selection levels even at high event rates without
compromising with the physical content of the data. 

To test the highest sustainable event rate with the current
combination of hardware and software, a sequential trigger was fed to
the system. The average event rate that the DAQ was able to cope with
was measured to be $4.74 \pm 0.01$ kHz. 

The novel Czochralski type silicon detector was tested using the new
DAQ by replacing one of the older detectors with the Cz-module.
The efficiency of the detector was measured to be 95 \%, the
resolution about 10 $\mu$m, and the signal to noise ratio about 10.

\section{CONCLUSIONS}
The results show that it is possible to construct an efficient and
accurate Data Acquisition System for a challenging LHC-like
environment of high event rates, high number of channels, limited
processing time and limited storage capability. It has been shown that
it is possible to build and successfully operate a DAQ system
fulfilling these criteria using the C++ programming language and
object orientated techniques on a Linux platform by using
sophisticated yet fast selection techniques and optimization based on
software architecture.

\begin{acknowledgments}
The authors wish to thank Dragoslav Lazic and the CMS HCAL group for
their valuable technical support and excellent collaboration at the
H2 beamline at CERN. The authors also wish to thank Dr.~M.~Kirsch and
Struck Innovative Systeme GmbH for excellent customer support and
assistance in using their VME modules.
\end{acknowledgments}



\end{document}